\documentclass[]{spie}  %>>> use for US letter paper
\usepackage{amsmath,amsfonts,amssymb}
\usepackage{rotating}
\usepackage{graphicx}
\usepackage[colorlinks=true, allcolors=blue]{hyperref}
\usepackage[utf8]{inputenc}

\title{HARPS3 for a Roboticized Isaac Newton Telescope}

\author[a]{Samantha J. Thompson}
\author[a,c]{Didier Queloz}
\author[b]{Isabelle Baraffe}
\author[b]{Martyn Brake}
\author[h]{Andrey Dolgopolov}
\author[a]{Martin Fisher}
\author[c]{Michel Fleury}
\author[i]{Joost Geelhoed}
\author[a]{Richard Hall}
\author[e]{Jonay I. Gonz\'alez Hern\'andez}
\author[d]{Rik ter Horst}
\author[d]{Jan Kragt}
\author[d]{Ramon Navarro}
\author[b]{Tim Naylor}
\author[c]{Francesco Pepe}
\author[f]{Nikolai Piskunov}
\author[e]{Rafael Rebolo}
\author[d]{Louis Sander}
\author[c]{Damien Segransan}
\author[a]{Eugene Seneta}
\author[b]{David Sing}
\author[g]{Ignas Snellen}
\author[g,d]{Frans Snik}
\author[g]{Julien Spronck}
\author[f]{Eric Stempels}
\author[a]{Xiaowei Sun}
\author[e]{Samuel Santana Tschudi}
\author[a]{John Young}

\affil[a]{University of Cambridge, Astrophysics Group, Cavendish Laboratory, J J Thomson Avenue, Cambridge, UK}
\affil[b]{University of Exeter, Physics and Astronomy, Exeter, EX4 4QL, UK}
\affil[c]{Geneva University, Observatoire Astronomique, 51 ch. des Maillettes, Versoix, Switzerland}
\affil[d]{NOVA Optical Infrared Instrumentation Group at ASTRON, PO Box 2, Dwingeloo, The Netherlands}
\affil[e]{Instituto de Astrof{\'\i}sica de Canarias (IAC), E-38205 La Laguna, Tenerife, Spain}
\affil[f]{Uppsala University, Regementsvaegen 1, Uppsala, Sweden}
\affil[g]{Leiden Observatory, Leiden University, Postbus 9513, Leiden, The Netherlands}
\affil[h]{Crimean Astrophysical Observatory, Nauchnij, 298409 Crimea, Russia}
\affil[i]{S\&T corporation, Olof Palmestraat 14, Delft, The Netherlands}

\authorinfo{Further author information: (Send correspondence to S.J.Thompson.)\\S. J. Thompson: E-mail: sjthompson@cantab.net\\ D. Queloz (PI): E-mail: dq212@cam.ac.uk}

\begin{document} 
\maketitle

\begin{abstract}
We present a description of a new instrument development, HARPS3, planned to be installed on an upgraded and roboticized Isaac Newton Telescope by end-2018.  HARPS3 will be a high resolution (R $\simeq$ 115,000) echelle spectrograph with a wavelength range from 380--690 nm.  It is being built as part of the Terra Hunting Experiment – a future 10 year radial velocity measurement programme to discover Earth-like exoplanets.  The instrument design is based on the successful HARPS spectrograph on the 3.6m ESO telescope and HARPS-N on the TNG telescope.  The main changes to the design in HARPS3 will be: a customised fibre adapter at the Cassegrain focus providing a stabilised beam feed and on-sky fibre diameter $\approx 1.4$ arcsec, the implementation of a new continuous flow cryostat to keep the CCD temperature very stable, detailed characterisation of the HARPS3 CCD to map the effective pixel positions and thus provide an improved accuracy wavelength solution, an optimised integrated polarimeter and the instrument integrated into a robotic operation.  The robotic operation will optimise our programme which requires our target stars to be measured on a nightly basis.  We present an overview of the entire project, including a description of our anticipated robotic operation.
\end{abstract}


\section{Introduction}
\label{sec:intro}

The number of discovered exoplanets has been increasing rapidly since the first exoplanet \cite{mayor95} was detected in 1995.  Since the launch of Kepler \cite{kepler2010} in 2009 there has been an even faster discovery rate of exoplanets by means of the transit method and we have learnt that planets are common objects orbiting stars.  It is most likely an observational bias, but the bulk of exoplanets detected so far have orbital distances less than 1 AU, in contrast to our own Solar System, so we are still some way from having a comprehensive view of the full diversity of planetary systems. 

Despite this significant progress over the last two decades and the many exoplanetary systems identified, including about a dozen ``habitable zone'' exoplanets from the Kepler space mission, we have not yet detected an ``Earth Twin'', making it difficult to set our own Solar System in context. This may be a selection effect, or it may be that our Solar system’s configuration is rarer than expected. 

To try and address this problem we are preparing a ``Terra Hunting Experiment'' to search for Earth-mass planets in Earth-like orbits around our nearest Solar-like stars.  We will conduct this search via the radial-velocity (RV) method which uses time series of high resolution spectra of a star to search for varying Doppler shifts of spectral lines that indicate the presence of another body in orbit about the star.  To achieve this we will build a high resolution spectrograph, called HARPS3, and install it on an upgraded and roboticized Isaac Newton Telescope (INT) on La Palma, Canary Islands.  A very successful instrument for accurate RV measurements with proven long-term stability \cite{HARPS2006} is the High Accuracy Radial velocity Planet Searcher (HARPS) \cite{HARPS2004} at the ESO La Silla 3.6m telescope.  Like HARPS-N \cite{HARPSN2012} on the 3.6m Telescopio Nazionale Galileo (TNG), our instrument for the Terra Hunting Experiment, HARPS3, will also be based on the original HARPS design.  

For an observer monitoring our Sun by the RV method, the Earth induces a signal of amplitude 9 cm/s with a 1-year period; our goal is to detect a signal at the $\sim10$ cm/s level at periods of 100+ days.  The Sun or similar stars produce RV signals with amplitudes an order of magnitude more than this due to intrinsic activity (oscillations, granulation, active regions, magnetic cycles etc.) on their surface, however these activity signals have characteristics and timescales that should allow them to be separated from the planetary signals.  This is an extremely challenging aim, but evidence suggests that adopting an intensive sampling strategy \cite{meunier2013} which allows accurate tracking of stellar activity and also helps to avoid false signal detections (due to incomplete phase coverage) can help us achieve our goals.  The Terra Hunting Experiment\footnote{http://www.terrahunting.org} (THE) will run for 10 years and during that time we will take regular, nightly measurements of a selection of our nearest and brightest solar-like stars (G and K-type dwarfs), searching for Earth-mass, long-period exoplanets.  The plan is to measure the most promising candidates {\em every} night for as long as they are visible in the year and repeat every year up to the 10 year duration of the survey.  THE will thus acquire the most extensive spectroscopic time series possible over a ten-year period.

To ensure an efficient execution of this programme, HARPS3 will be built with a robotic operation in mind.  This paper will further describe the HARPS3 instrument, the anticipated telescope upgrade and an overview of the planned robotic operation.

HARPS3 will also provide the ING community access to a Northern-Hemisphere high accuracy RV instrument.  The optimised, robotic scheduling enabled by HARPS3 on an upgraded INT will allow other time critical studies or time-series measurements like ours, which requires a specific sampling rate, or any other programme that would benefit from a more flexible or non-traditional allocation of observing time.  The facility will be an invaluable asset for follow-up observations to TESS \cite{TESS2015}, CHEOPS \cite{CHEOPS2014} and PLATO \cite{PLATO2014} discoveries.

\section{Description of the HARPS3 instrument}

The HARPS3 instrument will be a fibre-fed, high resolution, high stability, echelle spectrograph.  It will be installed at the 2.5m Isaac Newton Telescope (INT) in La Palma in the Canary Islands.  The design will be a close-copy of the very successful HARPS (South and more recently North) \cite{HARPS2004,HARPSN2012}.  HARPS3, like its predecessors, will be high resolution (R $\simeq$ 115,000) and via two optical fibres will provide simultaneous measurement of the science source and a spectral calibration source allowing accurate RV measurements to be made.  The wavelength range is 380--690 nm; the blue limit allows the measurement of Ca-II lines that are used to monitor the stellar activity cycle.  The ability to measure polarisation will also provide an alternative and complementary way to monitor stellar activity.  The instrument broadly consists of two main sub-systems: (1) the main body of the spectrograph housed in the large Coudé room of the INT and (2) the Cassegrain fibre adapter system installed at the Cassegrain focus of the telescope.

The main body of the spectrograph is housed in a thermally stable room; increasing degrees of thermal control are provided by using a nested enclosures arrangement.  These thermal enclosures, called HTE1, HTE2 and HTE3, each increase the level of thermal stability by an order of magnitude with the innermost HTE3 being stable to $\pm 0.01$K.  Within HTE3 is a custom built vacuum vessel housing the main spectrograph optics and detector.

The Cassegrain fibre adapter is attached to the telescope at the Cassegrain focus.  This unit contains the atmospheric dispersion corrector (ADC), a tip/tilt system, the calibration light projection optics, the polarimeter and the fibre selector mechanism.  The calibration unit (providing a number of calibration light sources) will be housed in the large Coudé room of the Isaac Newton Telescope, within the outer, large thermal enclosure (HTE1); the light from here is fibre-fed to the Cassegrain fibre adapter.

The following sections describe the HARPS3 subsystems in more detail.

   \begin{figure} [t]
   \begin{center}
   \begin{tabular}{c} 
   \includegraphics[height=7cm]{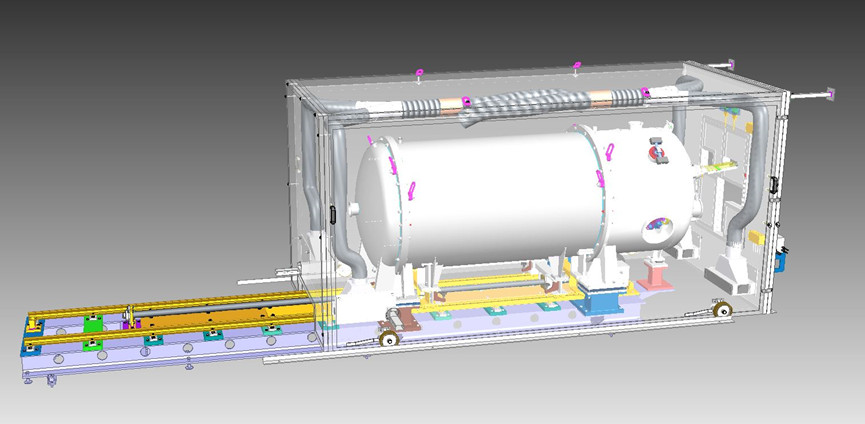}
   \end{tabular}
   \end{center}
   \caption[speccad] 
   { \label{fig:cad} 
The drawing shows the spectrograph system as installed within HTE2 (the middle thermal enclosure).  A rail system is embedded in the thermally insulated floor; the rails allow the vacuum vessel to be opened.  The vacuum vessel is surrounded by a thermal enclosure, HTE3 ($4 \times 2 \times 2$ metres) providing a higher degree of temperature control than HTE2.  The spectrograph optics and detector are housed within the vacuum vessel.  The cryostat (not shown) is located outside the vacuum vessel enclosure.  The length from the end of the rail (left) to the start of the vacuum vessel (right), not including the thermal enclosure, is $\sim6$ metres.}
   \end{figure}

\subsection{The Spectrograph}
The main body of the spectrograph will be installed in the INT large Coudé room.  The room will be fitted with suitable enclosures to provide a stable environment to operate the spectrograph; increasing degrees of thermal control are provided by using a nested enclosures arrangement with the innermost one being stable to $\pm 0.01$K.  The final enclosure in this arrangement is a custom built vacuum vessel housing the main spectrograph optics and detector.
The main components of the spectrograph consists of:
\begin{enumerate}
\item Vacuum system and mounting rails
\item Spectrograph optics and opto-mechanics
\item Detector system
\end{enumerate}

A CAD drawing of the spectrograph system, as it will be installed in the Coudé room is shown in Figure \ref{fig:cad}.  The design is a copy of that installed on HARPS-N except for some modifications made to the vacuum chamber to allow for the new continuous-flow cryostat.  The dimensions of the vacuum vessel (not including the rail system) are $1280 \times 1220 \times 3100$ mm (H$\times$W$\times$L) and the mass of the vessel (fully equipped with optics and detector) is approximately 2000 kg.

\subsubsection{Vacuum System}
The main spectrograph components are operated within a medium-level vacuum environment at around $5\times 10^{-3}$mbar inside pressure.  As shown in Figure \ref{fig:cad}, the vacuum vessel comprises three parts, the end-cap and main body of the vessel can be detached from the static module via a rail system.  The vacuum system minimises fluctuations in the instrument profile that can be brought about by changes in ambient air pressure; these would manifest as varying RV errors $> 1$ m/s.  An air-free operation for the spectrograph also removes any potential problems due to air turbulence which could also affect instrument stability.  The temperature control of the vacuum vessel is provided by the nested enclosures arrangement previously described.  Thermal gradients across the system can introduce significant RV errors, so it is important to maintain an even and constant temperature over the entire volume.  Numerous temperature sensors will be installed within the spectrograph volume and placed near key components to provide alerts should any significant temperature gradients develop during operation.

\subsubsection{Spectrograph Optics}

\begin{figure} [ht]
   \begin{center}
   \begin{tabular}{c} 
   \includegraphics[height=5cm]{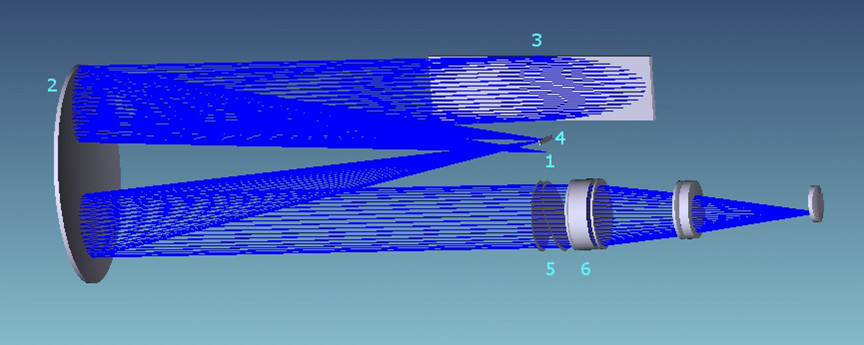}
   \end{tabular}
   \end{center}
   \caption[optics] 
   { \label{fig:optics} 
Ray trace of main spectrograph optical components ({\it design copy of HARPS-N}).  The light is injected at (1).  The large (770mm diameter) parabolic collimating mirror (2) is used in triple pass.  (3) Echelle grating, (4) flat fold mirror, (5) grism (cross-disperser) and from element (6) until the final focus are the 6 lenses of the camera objective.}
   \end{figure}

The optical ray trace showing the main spectrograph optics that are mounted and aligned within the vacuum vessel is given in Figure \ref{fig:optics}.  The optical design shown is that of HARPS-N \cite{HARPSN2012}, the original concept of which was based on that of UVES (the Ultraviolet and Visual Echelle Spectrograph on the VLT) \cite{UVES1992}.  It is an excellent solution for a relatively compact, cross-dispersed, very high resolution spectrograph using an R4 echelle grating.  The two optical fibres from the Cassegrain adapter provide the input to the spectrograph.  The input beam is converted to match the F-number of the collimator beam aperture.  A single large parabolic collimator is used in triple pass, rather than two separate mirrors, for reasons of increased stability.  Optimising stability is critical to performance given the desired RV accuracy for this instrument (a few 10’s cm/s).  The collimated beam is projected on to the echelle grating and the dispersed beam is reflected back on to the collimator; an intermediate spectrum is focused at the flat-folding mirror.  This spectrum is reflected back to the parabolic mirror for the third and last time, to collimate the beam for cross-dispersion by the grism.  The grism separates the superposed echelle orders.  The expected spectral resolution of the system is 115,000.

The camera objective focusses the images of the object and reference spectra, order by order, on the CCD detector.  The detector system contains a 4K x 4K back illuminated e2v CCD cooled by a continuous flow cryostat.

\subsubsection{Detector System}
HARPS3 will utilise the same large format, 4k $\times$ 4k e2V CCD that is used by HARPS-N.  It will be integrated into an improved design continuous-flow cryostat which minimises temperature fluctuations on the CCD for improved instrument stability.

Imperfections in the structure of the CCD can lead to a systematic error in the RV data, which could manifest as a false planet detection\cite{Dumusque15}.  Given our aim for an Earth-like planet detection and the extremely small RV signal this implies we will endeavour to perform a high accuracy mapping of the effective pixel positions of the HARPS3 CCD; see Hall et al. \cite{hall16} (also in this conference) for details.  Use of such a pixel map will provide an improved wavelength solution in the data reduction process.

\subsection{The Cassegrain Fibre Adapter System}
\begin{figure} [b]
   \begin{center}
   \begin{tabular}{c c}
   \includegraphics[height=8cm]{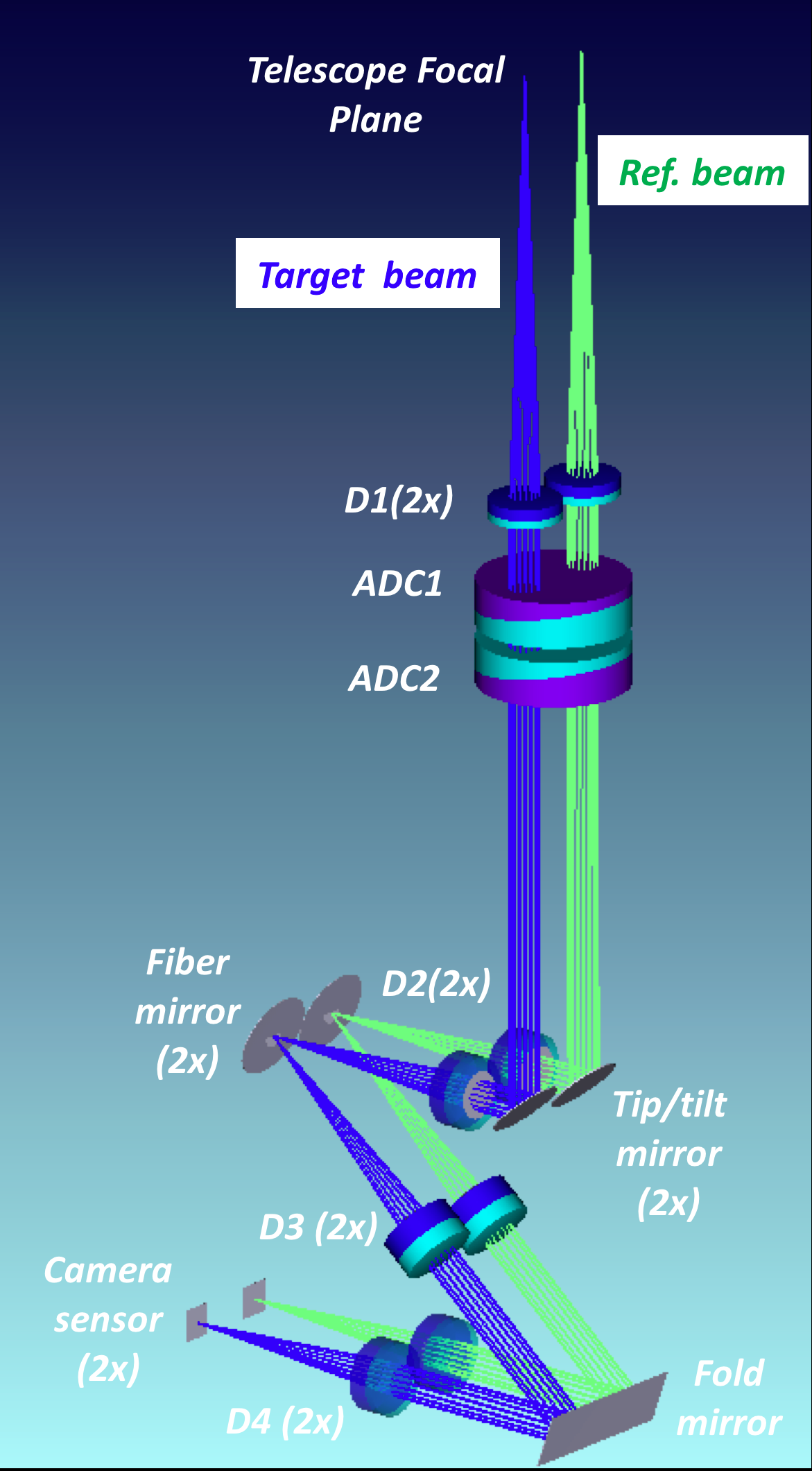}
   \includegraphics[height=8cm]{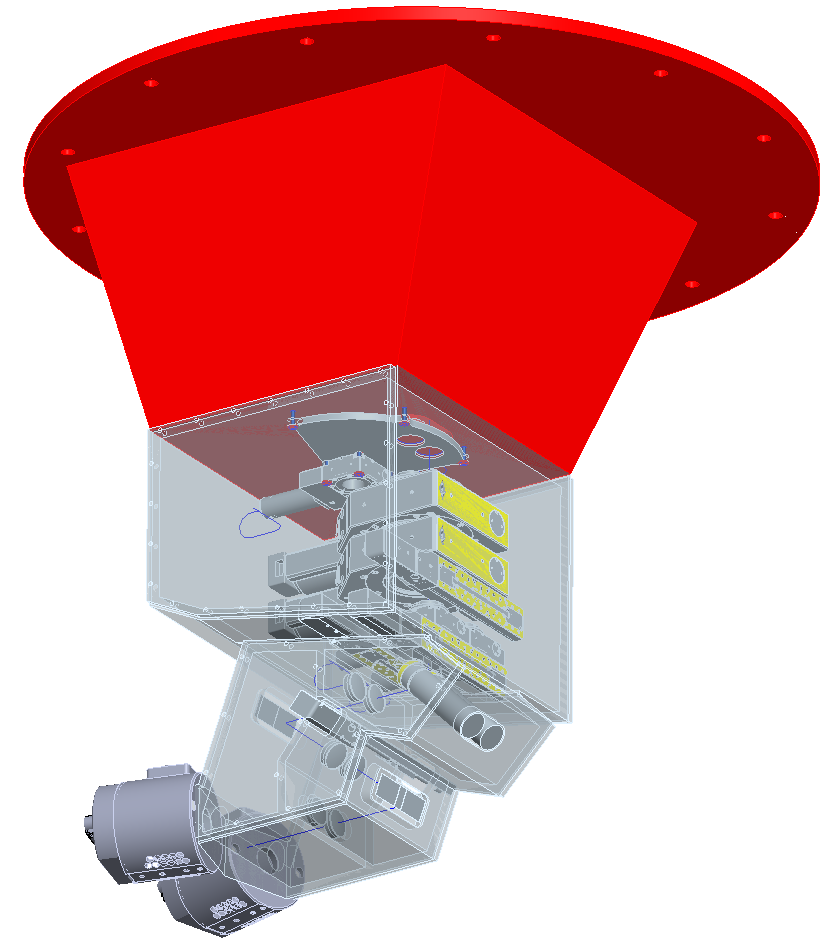}
   \end{tabular}
   \end{center}
   \caption[cass] 
   { \label{fig:cassunit} 
Drawings showing the current opto-mechanical design of the HARPS3 Cassegrain fibre adapter unit.  The optical view shown is for the target + sky reference mode.  For the standard observing mode for the Terra Hunting Experiment - target with simultaneous reference calibration lamp - the reference source does not pass through the ADCs.  The polarimeter optics are not shown in this view - they are inserted on a sliding mechanism after the ADCs (see Fig. \ref{fig:pol}).  The final design review is anticipated for the end of this year (2016).}
   \end{figure}

The main body of the fibre adapter is attached to the telescope at the Cassegrain focus.  It collects light from the science target and chosen calibration source and injects it into the two fibres that provide the feed into the spectrograph.  The light from the science target passes through an Atmospheric Dispersion Corrector (ADC) and is optimally positioned/stabilised on the fibre entrance via a tip-tilt mirror.  Figure \ref{fig:cassunit} shows the current optical and opto-mechanical design of the HARPS3 Cassegrain adapter unit.  Compared to the previous HARPSs designs, the optical paths to the two fibres are separate with each having its own tip/tilt mirror allowing optimal focal positioning on {\it both} fibres.  Due to the polarimeter in the system the tip-tilt mirrors are positioned after the polarisation modes have been separated.  The other change, due to the two separate beam design, is that the on-sky fibre separation is around $60\%$ larger than that in HARPS and HARPS-N ($\sim 160$ arcsec compared to $\sim 100$ arcsec); even for the worst case full-Moon scenario this is not a significant difference in terms of sky brightness effects.

Since the intent of this project is to preserve the design of the main body of the HARPS spectrograph as much as possible, the same type of 70 micron diameter fibres will be used to link between the Cassegrain and the spectrograph input.  These fibres have an octagonal cross-section to further improve the far and near-field scrambling effects.  Suitably designed optics in the fibre adapter unit take the f/15 telescope beam and convert it to the correct numerical aperture for the fibre input.  The scale change due to this gives an on-sky fibre diameter of approximately 1.4 arcseconds, which is well matched to the majority of site seeing conditions (see section \ref{site_cond}).

A polarimeter will be provided as an integral part of HARPS3 and will be a selectable option within the Cassegrain fibre adapter unit.  The HARPS3 polarimeter design will be based on the HARPS-POL \cite{harpspol2011} design --- both the INT and the ESO 3.6m telescopes are equatorial mount and both HARPS and HARPS3 use the straight-through Cassegrain focus.  There is also a polarimeter proposed for HARPS-N \cite{harpsnpol2014}, but it has the added complication of being on an alt-az telescope with an relatively unfavourable (for polarimetry) beam-feed from the Nasmyth focus.  Since HARPS3-POL will not be a retro-fit (as per HARPS-POL) it has fewer alignment issues and a reduced risk of polarisation cross-talk. 

\subsubsection{Guiding and Acquisition}
HARPS3 will have its own self-contained acquisition and guiding system.  This will allow accurate placement of the target star on the centre of the fibre entrance and maintain its position over the duration of an exposure.  The expected RV error due to guiding errors is proportional to $\delta \theta / \theta$, where $\delta \theta$ is the guiding error (in arcseconds) and $\theta$ is the on-sky fibre diameter (in arcseconds).  Our goal is to minimise any RV errors where possible, so our aim for the guiding accuracy is that the RV error contribution from this is $\sim 0.1 \times$ the photon limited RV error --- this has been plotted in Figure \ref{fig:photonerr}.

\begin{figure} [b]
   \begin{center}
   \begin{tabular}{c c}
   \includegraphics[height=8cm]{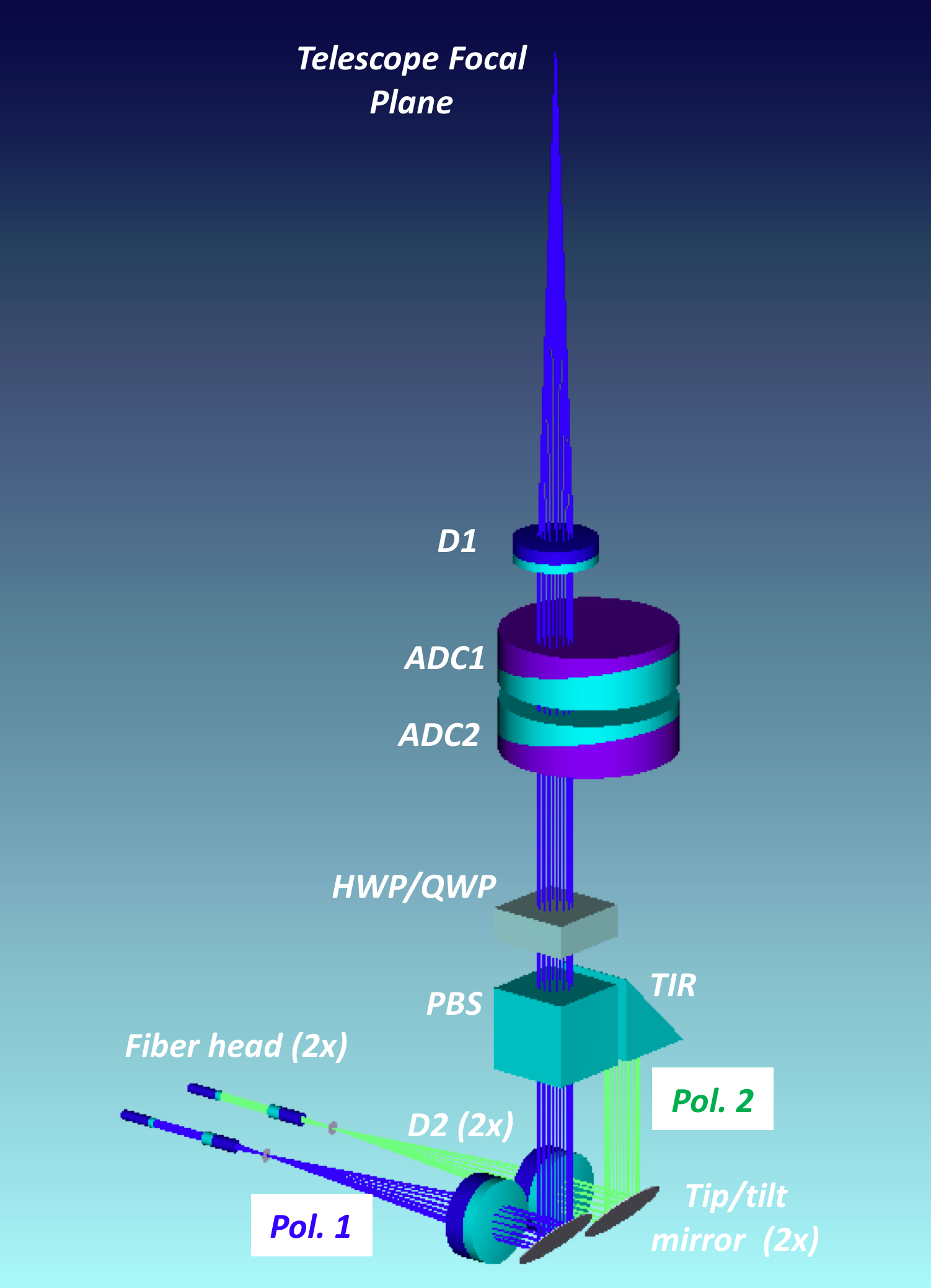}
   \includegraphics[height=8cm]{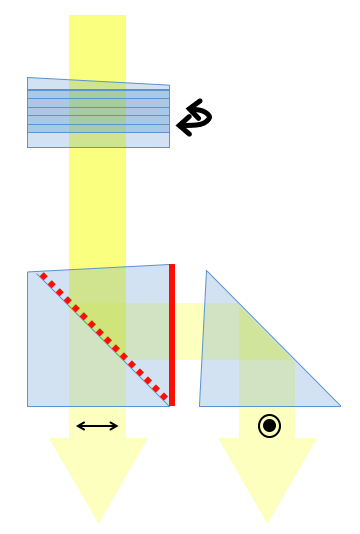}
   \end{tabular}
   \end{center}
   \caption[pol] 
   { \label{fig:pol} 
Drawings showing the current polarimeter mode design for HARPS3.  The diagram on the right shows a close-up of the polarimeter optical components - the ``superachromatic'' rotatable half or quarter wave plate, the polarising beam splitter cube (wire-grid type) and beam steering prism. }
   \end{figure}

\subsubsection{The HARPS3 Polarimeter}
The HARPS3 polarimeter can be automatically deployed in the optical path in the Cassegrain adapter to provide dual-beam spectro-polarimetry.  Figure \ref{fig:pol} presents an overview of the optical design for the polarimeter and the configuration of the Cassegrain adapter optics in polarimetry mode.  The polarimetric unit will enable the measurement of both circular (V/I) and linear polarization (Q/I, U/I).  The polarimetric sensitivity will be better than $10^{-4}$ after multi-line addition.  Such a level of sensitivity requires a time-modulated dual-beam system, in which two simultaneous spectra are measured for the same star but split according to perpendicular polarization. To ``switch the beams'' and cancel any differential effects due to transmission/aberration/flat-fielding effects, a step-wise rotatable wave-plate is installed above the polarizing beam-splitter (see right-hand diagram in Figure \ref{fig:pol}); the wave-plates are identical to the ones used in HARPS-POL.  Each polarized Stokes parameter (Q,U,V) is then obtained from 4 spectral intensity recordings: $2\times$ for the spatial splitting, and $2\times$ for the temporal modulation. These 4 measurements are then combined through a double-difference or double-ratio for the ultimate sensitivity without degradation by systematic effects. See Snik et al.\cite{snik2013} for further description of these strategies. The beam-splitter is based on wire-grid technology, which is now good enough to reach down into the deep blue \cite{bscube2003}.

As can be noted from the beam configuration in Figure \ref{fig:pol}, simultaneous reference source mode is disabled with the polarimeter in the beam path.

\subsubsection{The Calibration Unit}
The main body of the calibration unit will be housed in the Coudé room with the main body of the spectrograph.  The unit feeds the Cassegrain fibre adapter with a source of light used for calibration purposes.  The calibration light source is selectable and is provided via a number of different types of lamps or calibration sources (e.g. Thorium-Argon, Tungsten, Fabry-Perot).  HARPS3 will not initially be equipped with a laser frequency-comb calibration source, but it is an upgrade option if the resources become available in the future.

\subsection{The Fibre-link System}
The Cassegrain adapter is connected to the main body of the spectrograph via fibre optics.  This fibre link system consists of five key parts:
\begin{enumerate}
\item The fibre entrance head
\item The long octagonal fibre (telescope side)
\item The double scrambler unit
\item The short octagonal fibre (spectrograph side)
\item The fibre exit head
\end{enumerate}

The fibre entrance head converts the input beam into an F/4 beam for optimal fibre-injection.  The optics are precisely aligned, with the last micro-optic bonded to the front surface of the fibre to minimise coupling losses.  The fibres have an effective diameter of $70\mu$m and are of an octagonal cross-section.  As others have shown \cite{fibres2010,fibres2012}, a non-circular fibre cross-section provides excellent scrambling gains compared to circular fibres.  The double scrambler unit acts to further scramble the fibre-propagated light from the telescope by switching the near and far-field, but the lens system also provides a vacuum window interface into the vacuum vessel.  Further scrambling is then provided by a shorter length ($\sim 2$ meters) of the same octagonal fibre as the long length from the telescope (exact length to be decided).  The fibre exit head optics convert the light to the desired F-ratio of the spectrograph.

The two fibres in the fibre-link system which, coupled with the Cassegrain adapter system, allow for the following observing/calibration modes for HARPS3:
\begin{itemize}
\item Science target + calibration light source
\item Science target + sky
\item Spectro-polarimetry (full-Stokes) on science target
\item Calibration light source + calibration light source
\end{itemize}

\subsection{Software}
Compared to the previous HARPS instruments, HARPS3 plans to be operated in a largely robotic mode (no observers required on-site or remotely).  This requires certain key software elements to be rewritten with this new operation mode implicit in the architecture.  It also requires a slightly higher level of status reporting from the instrument control systems to ensure that if there are any problems the source can be identified and reset or investigated further.  In terms of robotic operations, there are two key pieces of new software – the observation (or master) control and an automated scheduler (previous schedulers have been manual).

\begin{sidewaysfigure} [p]
   \begin{center}
   \begin{tabular}{c} 
   \includegraphics[height=13
   cm]{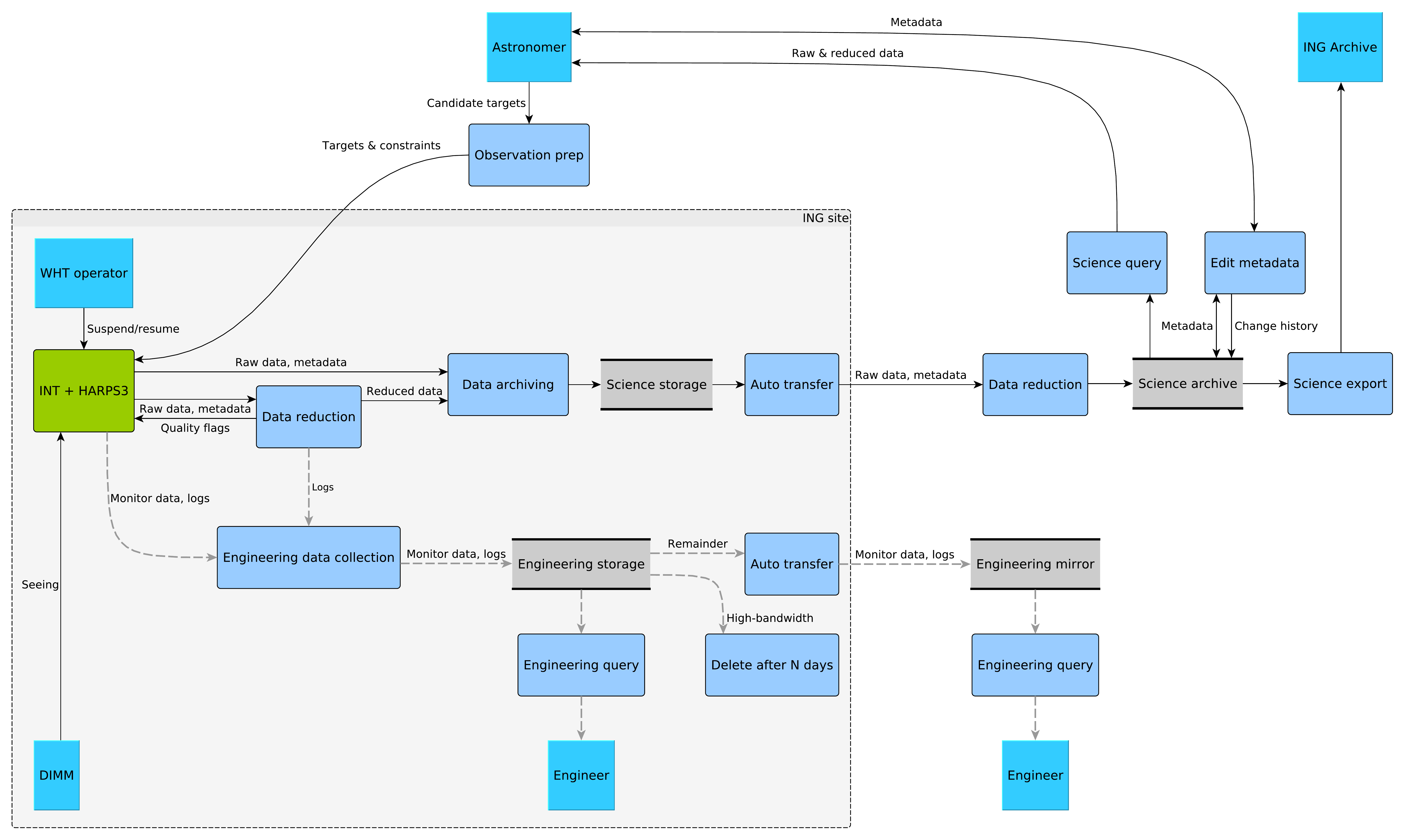}
   \end{tabular}
   \end{center}
   \caption[dataflow] 
   { \label{fig:dataflow} 
Contextual data flow diagram showing the data paths into and out of the user-facing HARPS3 software components. We anticipate that science data will be processed automatically at the observatory for quality control purposes. Calculation of the final science data products will be done off-site and the results and raw data stored in a science archive. To facilitate debugging of robotic operations the INT + HARPS3 system will capture detailed diagnostic data which can be reviewed by engineers and scientists both on- and off-site.}
   \end{sidewaysfigure}

A schematic of the HARPS3 dataflow is presented in Figure \ref{fig:dataflow}.

\begin{figure} [ht]
   \begin{center}
   \begin{tabular}{c} 
   \includegraphics[height=11cm]{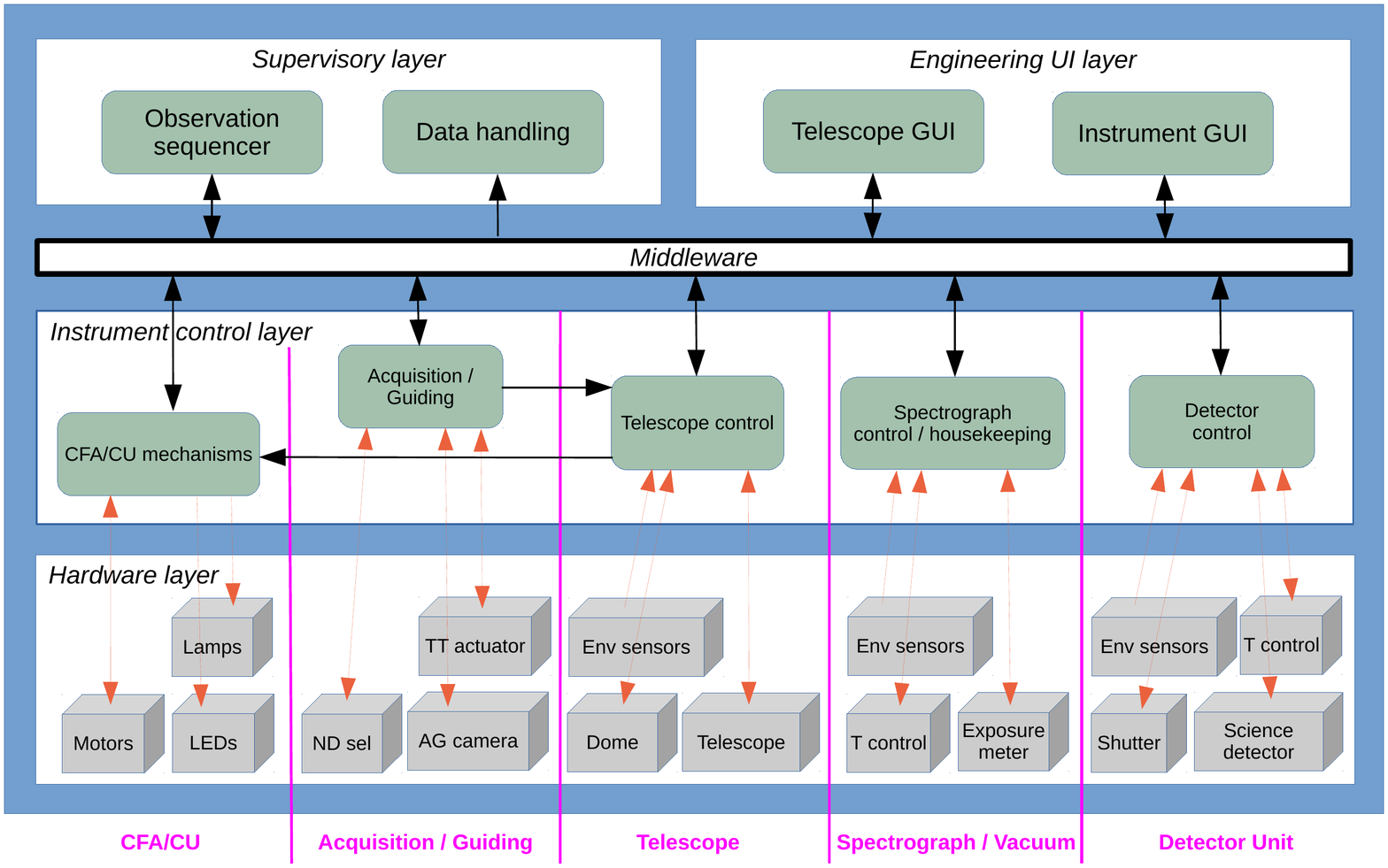}
   \end{tabular}
   \end{center}
   \caption[controlsys] 
   { \label{fig:controlsys} 
Architecture of the robotic control system for the INT and HARPS3. Science observations will be executed by the Observation Sequencer, which coordinates the actions of the sub-systems in the "instrument control layer". An automated scheduler (not shown) will decide which science observation to undertake next based on seeing estimates and other information, once the sequencer has completed (or abandoned) the current observation.}
   \end{figure}

An overview of the control system architecture is presented in Figure \ref{fig:controlsys}.

\subsection{Expected HARPS3 Performance}
As already described, the goal of the Terra Hunting Experiment is to detect Earth-like planets via the radial velocity method.  This implies that HARPS3 should be able to make measurements in which a small (down to 10 cm/s RV  amplitude), slowly varying (100--300 day period) signal can be detected (and repeatedly detected over several years). 

The measured performance of HARPS-N \cite{HARPSN2014} demonstrates a long-term RV accuracy at the $\sim 30$ cm/s level.  With this as our baseline design combined with the stability enhancements described and our daily data sampling strategy we hope to achieve a stable instrument accuracy that will meet our science goals.

\begin{figure} [ht]
   \begin{center}
   \begin{tabular}{c} 
   \includegraphics[height=7cm]{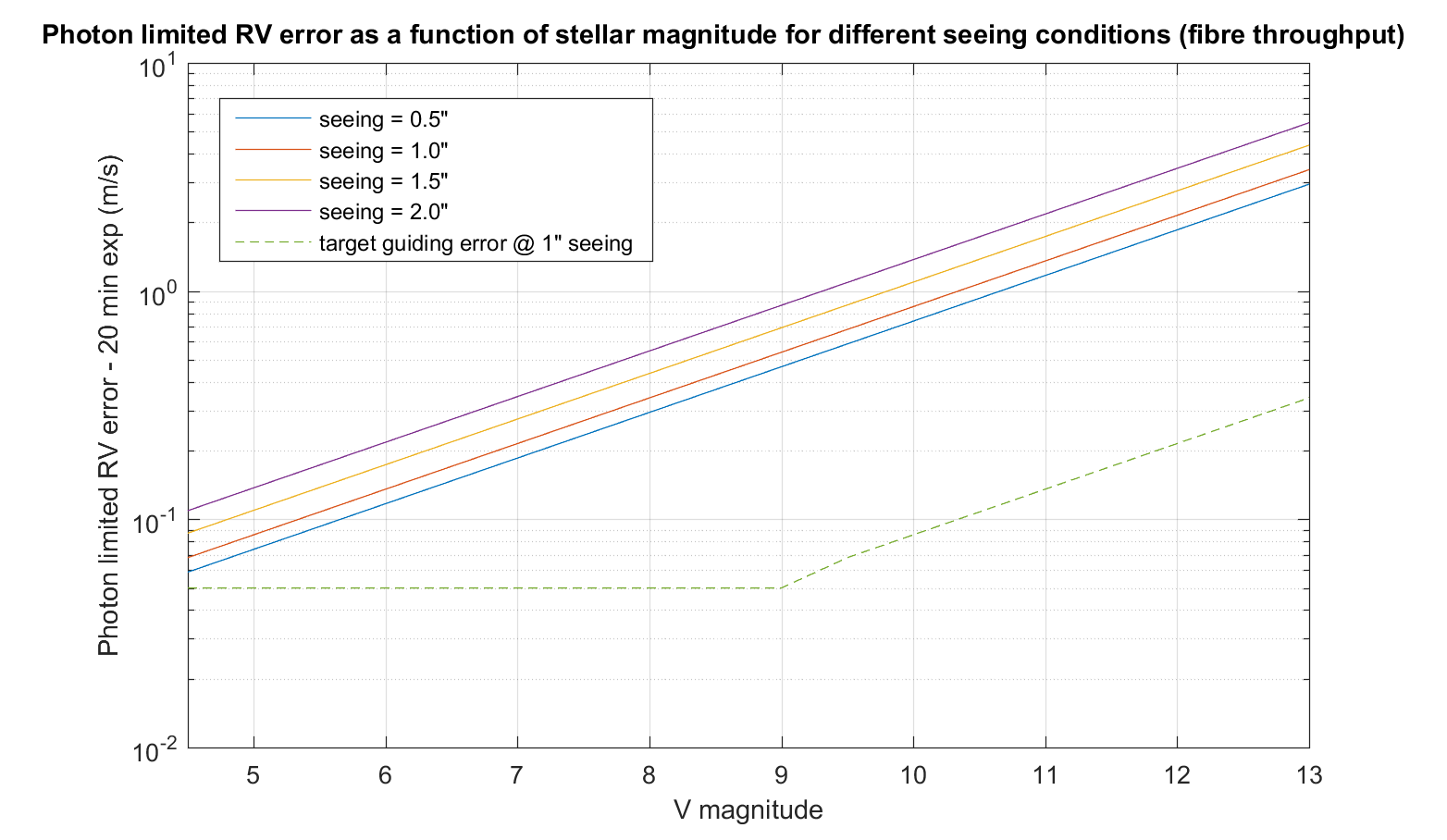}
   \end{tabular}
   \end{center}
   \caption[photonerr] 
   { \label{fig:photonerr} 
Graph showing how the photon limited RV errors change as a function of stellar V magnitude and the FWHM seeing size.  For these plots an exposure time of 20 minutes on target and average system efficiency of 5\% was assumed.  Changes in seeing conditions affect the throughput of the system since the on-sky fibre diameter $\approx 1.4$ arcsec.  The guiding error has a baseline of 0.05 arcsec RMS over an exposure but crosses into the regime of $0.1 \times RVerror_{photon}$ for fainter stars; the exact function for fainter stars is still under investigation.}
   \end{figure}
   
An estimate of the photon limited RV error as a function of stellar magnitude is given in Figure \ref{fig:photonerr}.  These calculations assume a typical exposure time for THE of 20 minutes and a cautious 5\% average system efficiency and are based on the theory described by Bouchy et al. (2001) \cite{bouchy2001}

\begin{figure} [ht]
   \begin{center}
   \begin{tabular}{c} 
   \includegraphics[height=6.5cm]{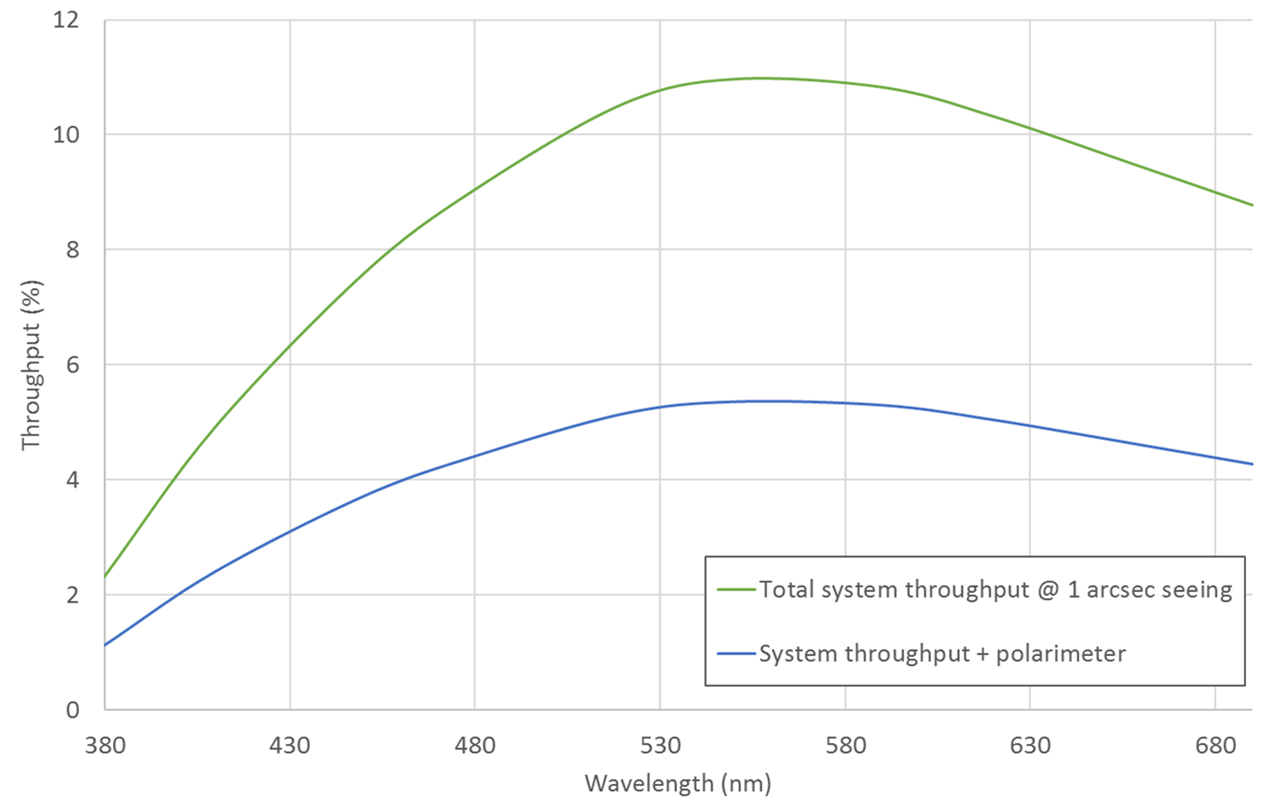}
   \end{tabular}
   \end{center}
   \caption[throughput] 
   { \label{fig:throughput} 
Graph showing the current expected system throughput as a function of wavelength.  The calculations for this plot include the transmission of the atmosphere.  The fibre entrance losses are calculated based on an assumed 1.0 arcsecond seeing condition (on-sky fibre diameter $\approx 1.4$ arcsec).  The total fibre length is not fully determined at this time, but is assumed to be 20 metres for these calculations.}
   \end{figure}

A preliminary estimate of the system efficiency is given in Figure \ref{fig:throughput}, both with and without the polarimeter deployed in the beam.  The length of fibre from the telescope to the spectrograph can significantly affect these results.  The minimum path from the telescope to the spectrograph is $\sim 15$ meters.  These calculations assume 20 meters in total for the optical fibre, but until the exact routing of the fibre cable is decided this number is open to change.  

\section{Refurbishment of the Isaac Newton Telescope}
The Isaac Newton Telescope (INT) \cite{intweb} is an equatorial-mount telescope in the Northern hemisphere on the island of La Palma in the Canaries.  The primary mirror is 2.54 m in diameter with a central obscuration of 0.914 m diameter; the effective collecting area is therefore 4.41 m$^{2}$.  At Cassegrain, the focal ratio is f/15.  The INT has two Coudé rooms, providing a minimum fibre run of $\sim 15$ metres from the Cassegrain focus.  Minimising the fibre link distance reduces light transmission losses, especially in the blue.

\subsection{Site Conditions}
\label{site_cond}
The median measured seeing on the INT is $\sim 1.3$ arcsec.  This is in contrast to the external local DIMM which measures $\sim 0.8$ arcsec.  It is likely that this difference is due to sub-optimal tracking/autoguiding which adds an extra $\sim 0.2$ arcsec; dome seeing is likely also adding an extra $\sim 0.3$ arcsec.  During the refurbishment of the INT improvements to the tracking performance will be sought.  Additionally we plan to improve the ventilation system in the main dome space which we hope will improve the dome seeing.  If these improvements are successful then it is believed that median seeing $< 1.0$ arcsec could be achieved.  

Six years of archived seeing data has been analysed from the ING RoboDIMM \cite{robodimm2003}.  Using all seeing data above an altitude of 60 degrees, 80\% of these data occur at seeing below 1.23 arcsec.  Using the lowest altitude zone (60--70 degrees), the 80\% limit lies at 1.34 arcsec.  This is a favourable result when using a fibre diameter of 1.4 arcsec.

On average, we can expect $\sim 20 \%$ time lost due to bad weather over the course of a year.  This is weighted towards the winter months and the distribution of time will also be different, i.e. in winter it is more likely to get longer downtime due to storms etc.  Such conditions will be factored into our programme simulations and strategies developed to mitigate the effects of this.

\subsection{INT Upgrade to Robotic Operations}
The nightly sampling rate and extended duration (10+ years) of the Terra Hunting Experiment requires an optimized operation strategy.  Ten years of manned operations does not appear to be the most efficient way of proceeding.  Our consortium plans to operate HARPS3 and the INT almost completely robotically.  Scheduling will be automated, with the telescope system adapting the programme to observing conditions.  Reliable and maintainable operations over this timescale are also essential.  

The INT and its systems have been assessed from an engineering perspective for its scope to become robotic (new TCS, automated dome controls + weather station) and we have concluded that this is a feasible upgrade.  The refurbishment is estimated to take 6-8 months to complete during which time the telescope will be non-operational.

The refurbishment involves removing all the current electronic systems, drive motors and controllers, servo amplifiers and push-button controls and replacing them with modern-day equivalents.  The interface could be a graphical touch-screen and programmable logic controller that can be remotely controlled if required.

With these refurbishments completed, the telescope can be operated in three main ways: (1) interactively (at the console), (2) robotically or (3) remotely.  In terms of operations for the survey programme, no observer will need to be present at the telescope and changes to the target list database can be uploaded remotely (to then be picked up by the automated scheduler).  A remote intervention may be required to address any system alerts.  It is anticipated that a duty engineer will be required to perform regular checks of the facility and change out empty liquid nitrogen tanks ($\sim$ once per week).

In terms of the risk of a robotic operation: although the INT is planned to be robotic, we have the advantage of ING staff available on site every observable night to fix small problems, and to ensure the safety of the telescope.  We anticipate that the INT will request permission from the Observing Support Assistant at the WHT before opening, they will be able to shut it down remotely should the need arise, and check it has closed itself in the morning.  Such a level of operation is less risky than that of telescopes where no immediate help is available.  Furthermore, risk associated with robotisation is considerably smaller than it would have been ten years ago.  Not only are there many examples of telescopes designed to be fully automated, retrofitting is now commonplace and we will benefit from using software systems used on other telescopes.

\section{Project organisation and Current Status}

HARPS3 and the Terra Hunting Experiment consortium comprises the following institutes: University of Cambridge, UK (PI: D. Queloz), University of Exeter, UK (Co-PI: I. Baraffe), Geneva University, Switzerland (Co-PI: F. Pepe), Institute of Astrophyics of the Canaries (IAC), Spain (Co-PI: R. Rebolo), The Netherlands Research School for Astronomy (NOVA), The Netherlands (Co-PI: I. Snellen) and Uppsala University, Sweden (Co-PI: N. Piskunov).  The Isaac Newton Group (ING) is also a member of the consortium management board to oversee the HARPS3 project and additionally assist in the upgrade of the telescope.

Given the current schedule, the call for tenders for the long lead-time, large optical elements (collimator, gratings, camera) of the spectrograph will go out this summer.

A year of temperature surveying at the INT will be complete by the end of this summer and the design of the temperature enclosures in the large Coudé room can then commence.

A detailed engineering assessment of the telescope was performed in the last quarter of last year (2015).  The call for tenders for the refurbishment of the INT is planned to be released before the end of summer this year; the actual refurbishment works are currently anticipated for end-2017/start-2018. 

The optical design of the Cassegrain adapter is at a level of stability to allow the polarimeter optical components to be ordered by early this summer.  The full Cassegrain adapter design should be complete by end-2016.

With the current schedule, first-light for HARPS3 on a refurbished Isaac Newton Telescope is planned for end-2018.

\acknowledgments % equivalent to \section*{ACKNOWLEDGMENTS}       
 
R.H. acknowledges the Science and Technologies Facilities Council (STFC) for his PhD studentship award (2015). \\
J.I.G.H. acknowledges financial support from the Spanish Ministry of Economy and Competitiveness (MINECO) under the 2013 Ram\'on y Cajal program MINECO RYC-2013-14875.\\ 
J.I.G.H., R.R., and S.S.T. also acknowledge the Spanish ministry project MINECO AYA2014-56359-P.\\
NP and ES are grateful to Knut and Alice Wallenberg Foundation for a generous support of the Swedish contribution to the THE project.\\ 
AD acknowledges the support from Russian Foundation for Basic Research as part of research grant 15-52-12371.\\

% References
\bibliography{report} % bibliography data in report.bib
\bibliographystyle{spiebib} % makes bibtex use spiebib.bst

\end{document}